# Surface Reconstruction with Higher-Order Smoothness

Rongjiang Pan[1], Vaclav Skala[2]

[1]*School of Computer Science and Technology, Shandong University, Jinan, China, 250100,*
*panrj@sdu.edu.cn*
[2]*Centre of Computer Graphics and Data Visualization, Department of Computer Science and Engineering, University of West Bohemia, Plzen, Czech Republic*

## Abstract

*This work proposes a method to reconstruct surfaces with higher-order smoothness from noisy 3D measurements. The reconstructed surface is implicitly represented by the zero level-set of a continuous valued embedding function. The key idea is to find a function whose higher-order derivatives are regularized and whose gradient is best aligned with a vector field defined by the input point set. In contrast to methods based on the first-order variation of the function that are biased towards the constant functions and treat the extraction of the isosurface without aliasing artifacts as an afterthought, we impose higher-order smoothness directly on the embedding function. After solving a convex optimization problem with a multi-scale iterative scheme, a triangulated surface can be extracted using the marching cubes algorithm. We demonstrated the proposed method on several data sets obtained from raw laser-scanners and multi-view stereo approaches. Experimental results confirm that our approach allows us to reconstruct smooth surfaces from points in the presence of noise, outliers, large missing parts and very coarse orientation information.*

**Keywords**: Surface reconstruction, Higher-order smoothness, Convex optimization

## 1 Introduction

Reconstructing three-dimensional digital models from real world objects is one of the major research topics in computer graphics as well as in computer vision. The majority of the developed geometric acquisition techniques, such as active and passive range sensing, usually measure a large number of 3D points. However, the discrete points are not useful for many practical applications although point-based geometry representation has been proposed [1]. Thus, reconstructing watertight surfaces from a set of sparse points is becoming a common step in the acquisition process. The problem has been researched extensively and many techniques have been developed over the past two decades [2]-[11]. However, surface reconstruction remains a difficult and, in general, an ill-posed problem since noise and outliers often contaminate the scanned data. Moreover, due to inaccessibility during scanning and some material properties, there will be cases where points are missing or incomplete.

To cope with most of the deficiencies, energy-based methods, which combine the quality of fit to data with surface regularization, are particularly appropriate for robustly constructing surfaces from sampled point sets. Recently, global optimization frameworks, e.g. graph-cut [12] and convex relaxation techniques [13], have been applied to the surface fitting problem, where the surfaces are represented implicitly by the binary-valued indicator functions. The binary volume techniques focus on segmenting a voxel as the interior or the exterior of the underlying shape. Once the function is computed, a triangulated surface model can be efficiently recovered using an isosurface extraction algorithm such as marching cubes [14]. Nevertheless, the isosurfaces often suffer from aliasing artifacts and require post-processing to achieve smooth surfaces [15].

In this paper, we propose to impose higher-order smoothness directly on a continuous-valued embedding function. Moreover, instead of measuring the distance between the surface and the given noisy data points, we wish to compute the function whose gradient is best aligned with an estimated coarse normal field. As a result, the surface reconstruction problem is formulated as a convex optimization, whose minimum yields higher quality surfaces. Computationally, the function is discretized on a regular 3D grid and constructed by solving a large sparse linear system using a multi-scale iterative scheme.

The paper is organized as follows. We give some related work in the next section. Section 3 presents our energy formulation and the implementation details are provided in Section 4. In Section 5, we show some experimental results and a brief summary is concluded in Section 6.

## 2 Related Work

In the presence of noise and inhomogeneous sample density, most popular approaches for surface reconstruction fit continuous valued or characteristic (inside-outside) functions to the input point set and then extract the reconstructed surface as an appropriate isosurface of this function. The pioneered work by Hoppe [2] defines the implicit function as the signed distance to the tangent plane of the closest input point. Signed distance can also be merged together using an averaging process into a volumetric grid function [3]. Since both methods did not employ any surface regularization, they are prone to problems when the data contains large gaps and outliers. A Markov Random Field (MRF) based regularization is recently applied to the signed distance field [16]. Several methods define the implicit function as the weighted sum of radial basis functions (RBFs) centered at some points [6]-[9]. The fitting is done by solving a large linear system, where the trivial solution and the computational problem in practical solutions ought to be overcome. An alternative approach combines several local distance functions over an octree structure using a multilevel partition-of-unity [10]. The moving least squares (MLS) method [30] locally approximates the surface with polynomials. Issues of these methods are the lack of robustness of the local approximations and the presence of spurious surface artifacts. Poisson surface reconstruction [11] computes a smoothed indicator function (defined as 1 at points inside the model and 0 at points outside) over an octree using the Poisson equation. The gradients for this function approximate a vector field defined by the sample points. The method indirectly minimizes the membrane energy of the clipped signed distance field and this is not always optimal [16]. Most of these methods are sensitive to the accuracy of point orientations and varying sampling density. To handle noisy, incomplete and uncertain data statistical methods are also applied in the surface reconstruction domain [31]. However, they suffer from the need of user parameters and are relatively slow.

For robust surface fitting, discrete graph-cut [12] or continuous convex relaxation [13] schemes have recently been used where surfaces are represented implicitly by the binary characteristic function. The methods consider the problem of surface reconstruction as a three-dimensional segmentation task and employ total variation constraints. However, isosurfaces extracted from binary segmentation of discrete grids often exhibit aliasing artifacts and require post-processing steps [15].

In implicit surface frameworks, a lack of accurate information about the surface orientation at the point samples is known to be a main challenge. The generation of the implicit function relies on a way to distinguish between the inside and outside of the closed surface. Various methods have been proposed to obtain orientation information, such as estimating point normals using local principal component analysis (PCA) [2], classifying poles of the Voronoi diagram of the input points [17], heuristically computing inside/outside constraints [18]. In the presence of noise or thin features, the additional information is highly unreliable and often leads to an erroneous surface reconstruction. Some approaches try to reconstruct a surface approximation from unoriented point sets [19][20][21].Without orientation information, however, these algorithms may lead to over-smoothing surface [19], cannot fill large gaps [20] or deal with large data sets [17] [21].

# 3 Our Approach

## 3.1 Problem Formulation

Let $S$ be a set of sampled data points lying on or near the surface $\partial M$ of an unknown three-dimensional model $M$ . Each sample $s \in S$ consists of a point $p$ and a weak estimate of global surface orientation $\mathbf{n}$ . We wish to construct a continuous scalar-valued function $f(p)$ , $p \in \Omega$ defined over a closed and bounded domain $\Omega \subset \mathbf{R}^3$ , whose isosurface is the best fitting of the data points. A triangulated surface $\partial \tilde{M}$ can then be reconstructed by extracting the corresponding isosurface from the computed function $f$ .

As a smooth function $f$ leads to a smooth isosurface, we directly impose the smoothness on $f$ so that the reconstructed surface $\partial \tilde{M}$ possesses a certain degree of smoothness. The idea is extensively used in image segmentation [22][23].

## 3.2 Measure of Smoothness

In order to measure the energy or smoothness of a function, we can define a norm on the solution space: functions with a small norm are smoother than those with a large norm.

Membrane energy is used in the Poisson surface reconstruction algorithm [11] and constrained FEM reconstruction [18],

$$\mathrm{E}_s = \iiint_\Omega \left| \nabla f(x, y, z) \right|^2 dx dy dz = \iiint_\Omega \left[ f_x^2(x, y, z) + f_y^2(x, y, z) + f_z^2(x, y, z) \right] dx dy dz \qquad (1)$$

, where the subscripts denote differentiation and $\left|\nabla f\right|$ is the Euclidean norm on $\mathbf{R}^3$. It is more common to use the $L_1$ norm, which is often called total variation [22],

$$\mathrm{E}_s = \iiint_\Omega \left|\nabla f(x,y,z)\right| dxdydz = \iiint_\Omega \left[ f_x^2(x,y,z) + f_y^2(x,y,z) + f_z^2(x,y,z) \right]^{1/2} dxdydz \qquad (2)$$

The methods based on some function of the first-order variation $\left|\nabla f\right|$ are biased towards the constant functions, as they are the globally optimal functions under this measure.

To achieve higher-order smoothness, we suggest using the measure that integrates the squared second derivates:

$$\mathrm{E}_s = \iiint_\Omega \left[ f_{xx}^2(x,y,z) + f_{yy}^2(x,y,z) + f_{zz}^2(x,y,z) \right] dxdydz \qquad (3)$$

so that the globally minimal functions are the polynomials of degree at most three. Thus, the isosurfaces include all planar surfaces and some quadric surfaces. Alternatively, when the mixed derivatives are included, the measure

$$\mathrm{E}_s = \iiint_\Omega \left[ f_{xx}^2(x,y,z) + f_{yy}^2(x,y,z) + f_{zz}^2(x,y,z) + 2f_{xy}^2(x,y,z) + 2f_{xz}^2(x,y,z) + 2f_{yz}^2(x,y,z) \right] dxdydz \quad (4)$$

becomes rotationally invariant and biases the function towards a linear polynomial whose isosurfaces include all planar surfaces.

### 3.3 Data Fitting

Accounting for the uncertainties in the input data points, flux-based functionals are well justified data fit measures and are less sensitive to the orientation errors, as demonstrated in [12]. Moreover, nearly all capture devices can provide some kind of point normal information. For example, directions towards the sensor are often known for data points. From the orientations for the data points, a vector field $F : \Omega \to R^3$ encoding the normal directions is estimated by a smoothing filter. We then wish the gradient of the implicit function is best aligned with the vector field. A reasonable data fitting measure is the integration of the dot product between the gradient field $\nabla f$ and the vector field $F$ over the domain $\Omega$:

$$\mathrm{E}_d = \int_\Omega < \nabla f, F > \qquad (5)$$

The more similar between the two functions in the domain $\Omega$, the larger the measure is. Using the integration by parts, we can derive the following equivalent data energy:

$$\mathrm{E}_d = -\int_\Omega f \cdot \mathrm{div} F \qquad (6)$$

where $\mathrm{div} F$ is the vector field's divergence.

### 3.4 Energy Formulation

To obtain a global energy that can be minimized, the smoothness and data energy are combined together,

$$\mathrm{E} = \lambda \mathrm{E}_s \text{-} \mathrm{E}_d \qquad (7)$$

where the parameter $\lambda > 0$ controls how smooth the solution should be and determines in some sense the smallest feature that will be maintained in the reconstructed surface.

The functional in (7) is convex because each of the terms is a convex one. The gradient and Laplacian in the first one are linear operators and the second one is linear. To make the global minima well defined, we can simply constrain the solution to lie in a fixed interval, e.g. [-1,1] for all $p \in \Omega$.

## 4 Implementation

We now describe the implementation details of the proposed approach. In order to find the minimum of the continuous problem, the function $f(p)$ and the vector field $F$ are discretized on a regular 3D grid. For derivative and divergence operators, the judicious selection of the uniform space division results in simple discrete forms.

Moreover, when a more advanced structure is used, e.g. the octree, areas with no samples are represented by very big cells and have low resolutions in the final surface, as shown in Section 5.

## 4.1 Vector Field Estimation

The vector field is computed using a method similar to the one described in [12]. Initially, the weak estimate of global surface orientation $\mathbf{n}$ at each sample $s \in S$ is distributed to its eight nearest grid vertices as follows: $\mathbf{n} \cdot (1-dx)(1-dy)(1-dz)$ , $\mathbf{n} \cdot dx(1-dy)(1-dz)$ , $\mathbf{n} \cdot (1-dx)dy(1-dz)$ , $\mathbf{n} \cdot (1-dx)(1-dy)dz$ , $\mathbf{n} \cdot dx(1-dy)dz$ , $\mathbf{n} \cdot (1-dx)dydz$ , $\mathbf{n} \cdot dxdy(1-dz)$ , $\mathbf{n} \cdot dxdydz$ , where $dx, dy$ and $dz$ are the differences between the coordinates of point $p$ and the smallest coordinates among the eight grid vertices divided by grid spacing $h$ . In order to approximate a dense vector field $\{F(p) \big| p \in \Omega\}$ , we smooth the vector field with a Gaussian. For efficiency, we approximate the Gaussian by the $n$ -th convolution of a box filter with itself:

$$B(t) = \begin{cases} 1 & |t| < h \\ 0 & otherwise \end{cases}$$

where $h$ is the size of grid cell and we choose $n = 3$ in our implementation. Then, $\text{div}F = \dfrac{\partial F_x}{\partial x} + \dfrac{\partial F_y}{\partial y} + \dfrac{\partial F_z}{\partial z}$ is approximated by standard central differences where $F = \begin{pmatrix} F_x & F_y & F_z \end{pmatrix}^{\mathrm{T}}$ .

## 4.2 Optimization

After discretization, the corresponding data energy in (6) become

$$\mathrm{E}_d = -\sum_{i,j,k} \left[ f(i,j,k) \cdot \text{div}F(i,j,k) \right] \tag{8}$$

and the derivatives in (4) can be approximated as

$$f_{xx} \approx \frac{1}{h^2}[f(i-1,j,k) - 2f(i,j,k) + f(i+1,j,k)]$$

$$f_{yy} \approx \frac{1}{h^2}[f(i,j-1,k) - 2f(i,j,k) + f(i,j+1,k)]$$

$$f_{zz} \approx \frac{1}{h^2}[f(i,j,k-1) - 2f(i,j,k) + f(i,j,k+1)]$$

$$f_{xy} \approx \frac{1}{4h^2}[f(i+1,j+1,k) - f(i+1,j-1,k) - f(i-1,j+1,k) + f(i-1,j-1,k)] \tag{9}$$

$$f_{xz} \approx \frac{1}{4h^2}[f(i+1,j,k+1) - f(i+1,j,k-1) - f(i-1,j,k+1) + f(i-1,j,k-1)]$$

$$f_{yz} \approx \frac{1}{4h^2}[f(i,j+1,k+1) - f(i,j+1,k-1) - f(i,j-1,k+1) + f(i,j-1,k-1)]$$

The total energy of the discretized problem can be written as a quadratic form

$$\mathrm{E} = \lambda \mathrm{E}_s - \mathrm{E}_d = \lambda \mathbf{x}^T \mathbf{A} \mathbf{x} + \mathbf{x}^T \mathbf{b} \tag{10}$$

where $\mathbf{x} = [f(0,0,0) \cdots f(i,j,k) \cdots f(m-1,n-1,l-1)]^{\mathrm{T}}$ and $\mathbf{b} = [\text{div}F(0,0,0) \cdots \text{div}F(i,j,k) \cdots \text{div}F(m-1,n-1,l-1)]^{\mathrm{T}}$ for the grid of resolution $m \times n \times l$ and $i = 0 \cdots m-1, j = 0 \cdots n-1, k = 0 \cdots l-1$ . Minimizing the quadratic form is equivalent to solving the sparse linear system

$$\mathbf{A}\mathbf{x} = -\frac{1}{2\lambda}\mathbf{b} \tag{11}$$

In our implementation, we used the Gauss-Seidel method for determining the solution of the linear system and started with an initialization $\mathbf{x}^0 = [0 \cdots 0]^T$. The iteration terminates when the change in successive iterates, $\left\| \mathbf{x}^{k+1} - \mathbf{x}^k \right\|$, reaches the precision given by the user.

### 4.3 Multi-scale Solver

Due to uniform space division, a multi-scale solver is easy to implement. We use three levels of grid resolution. The divergence is computed at finest grid only once, and down-sampled by summation. The results computed on a coarse grid are up-sampled as a good initialization on the next fine level. Therefore, the optimization may concentrate on a narrow band around the input points at fine levels.

### 4.4 Meshing

In order to reconstruct a triangulated surface $\partial \tilde{M}$, it is necessary to select an isovalue and extract the corresponding isosurface from the computed function $f$. The isovalue is selected as the average value of $f$ at the sample positions:

$$\gamma = \frac{1}{|S|} \sum_{s \in S} f(p) \tag{12}$$

where $f(p)$ denotes the trilinear interpolation to the eight nearest grid vertices of $s$. Finally, we extract the isosurface using an adaptation of the implementation code of Marching Cubes published by Lewiner [24].

## 5 Results

We validated our approach on a series of experiments. The proposed method was implemented in C++ and run on a notebook with 2.26GHz Core 2 Duo CPU and 2GB of RAM.

In order to compare the behavior of the different energy models, we began with an artificial data set used in [16]. The points are sampled randomly from a collection of primitives. Figure 1 clearly shows that the high order derivates in (3) and (4) result in a smoother surface than using the first-order variation. Table 1 gives the accuracy of the reconstructed models shown in Figure 1 in terms of the root mean square (RMS) of distances from the input points to the nearest points on the surface. The smoothness of the reconstructed surface is estimated by computing the mean curvature and Gaussian curvature using the trimesh2 library [29].

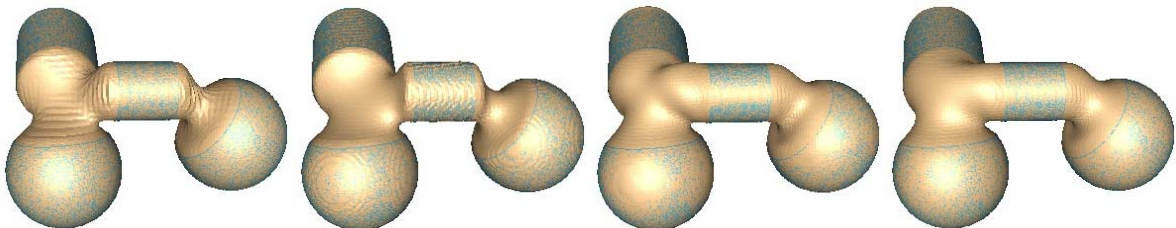

Figure 1: Surface reconstruction on an artificial data set. The points are used as input. From left to right: the results using the energy terms in (1), (2), (3) and (4).

| Energy model | Triangles | RMS | Average Mean curvature | Maximum Mean curvature | Average Gaussian curvature | Maximum Gaussian curvature |
|---|---|---|---|---|---|---|
| (1) | 49,524 | 0.14 | 0.0207517 | 0.778069 | 0.000452 | 10.03470 |
| (2) | 49,156 | 0.18 | 0.0228084 | 0.288104 | 0.000083 | 0.056263 |
| (3) | 51,732 | 0.15 | 0.0214885 | 0.127483 | 0.000128 | 0.017155 |
| (4) | 51,112 | 0.11 | 0.0212692 | 0.102347 | 0.000120 | 0.010869 |

Table 1: The energy model, the number of triangles in the reconstructed model, the RMS of distances from the input points to the nearest points on the surface, the average and maximum mean curvature, the average and maximum Gaussian curvature.

We also processed a number of range data sets from the Stanford 3D Scanning Repository [25]. To demonstrate the robustness of our method to orientation errors, the registered raw range scans were treated as a collection of 3D points and a single orientation vector corresponding to scan viewing direction was assigned to all points in the same scan. In the following, the results are based on the energy term in (4).

Figure 2 illustrates the effects of the parameter $\lambda$ for the Bunny model. It is observed that the parameter $\lambda$ affects the fitness to the sample points and smoothness of the surface. Large values of $\lambda$ lead to increased smoothing. In our experiments, we got reasonable results with $\lambda$ between 0.1 and 0.3.

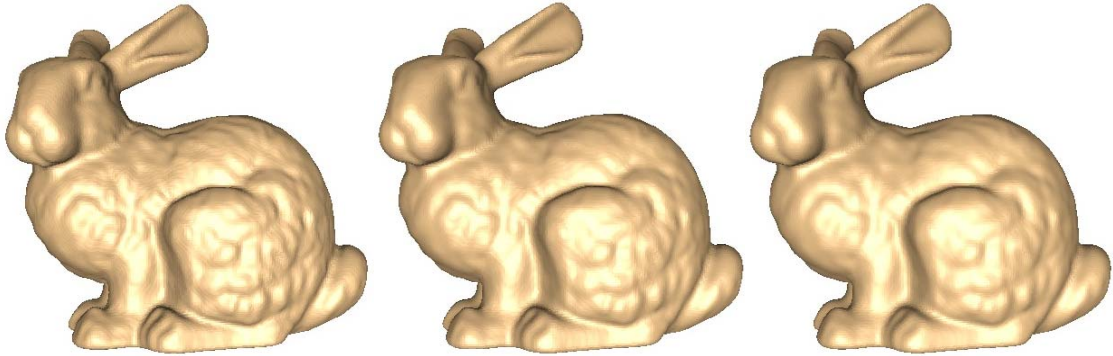

Figure 2: Reconstruction of the Bunny model at various values of $\lambda$ : 0.1 (left), 0.5 (middle), 1.0(right)

We compared the results of our method to the results obtained using Poisson surface reconstruction method [11] and the MRF algorithm [16]. Surfaces reconstructed with the three methods for the Dragon model are shown in Figure 3. It can clearly be seen that the Poisson method (default parameters and depth=10) and the MRF method (type=points with approximate normals ) were unable to handle the original raw scans with coarse orientation estimates (one direction towards the scanner per scan), while our method produced an reasonable result.

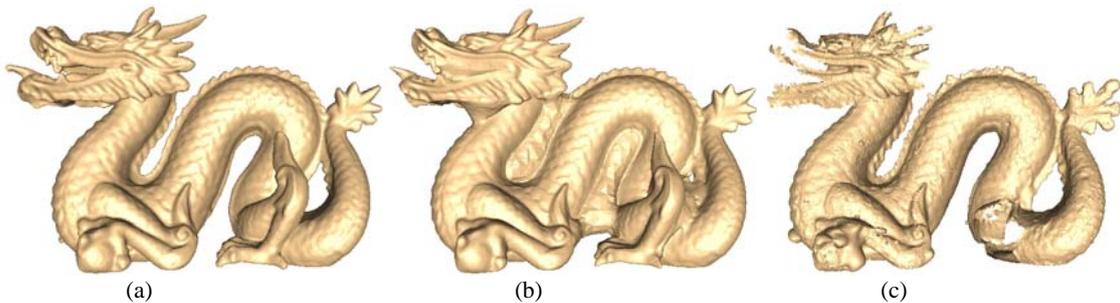

(a)                    (b)                    (c)

Figure 3: Comparison of three different algorithms. (a) The result of our method.  (b) The result of the Poisson surface reconstruction algorithm and (c) the result of the MRF algorithm.

To study scalability with large variations in sampling density and some outliers, we removed 98% of points from one-half of Armadillo and kept the outliers added by scanning process. Unlike [12] that used non-uniform Euclidean regularization, our method was able to handle the 50-to-1 difference in density and tolerate outliers without using any other information, as shown in Figure 4.

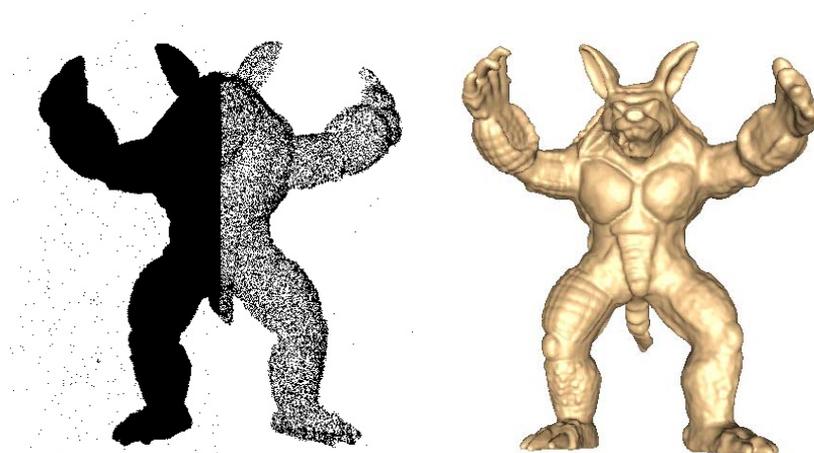

Figure 4: Reconstruction result (right) of the Armadillo range scans with 50-to-1 difference in density (left).

An input data set with several large holes is given in Figure 5. The hole-filling capabilities can be seen from the reconstructed surface.

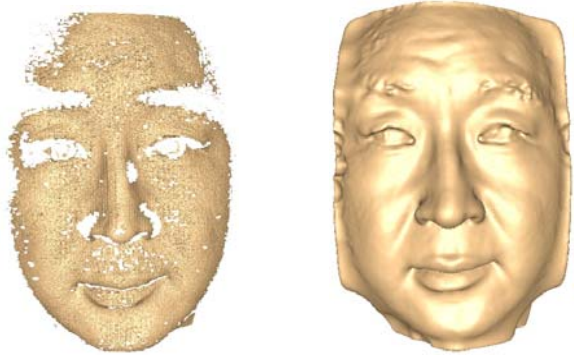

Figure 5: Reconstruction result (right) of an input data set with large holes (left).

The presented algorithm has also been tested on several data sets produced by multi-view stereo algorithm. After taking a collection of photographs with a digital camera, we estimated the camera parameters using the structure from motion software Bundler [26]. Then, a patch-based multi-view stereo algorithm [27] can produce a set of oriented points covering the surface of the object. The oriented points are used as input to our surface reconstruction method and the reconstructed surface is textured from the photographs using a modified version of the technique described in [28].

Figure 6 shows a 3D reconstruction of the Confucius statue from 40 photos. Because the statue is about three meters high, some parts of the model are not imaged and the obtained point set has several large holes. Our method fills the holes in a plausible way. For a 30cm high statue of the Goddess of Mercy, the result of our surface reconstruction method from 45 photos is shown in Figure 7.

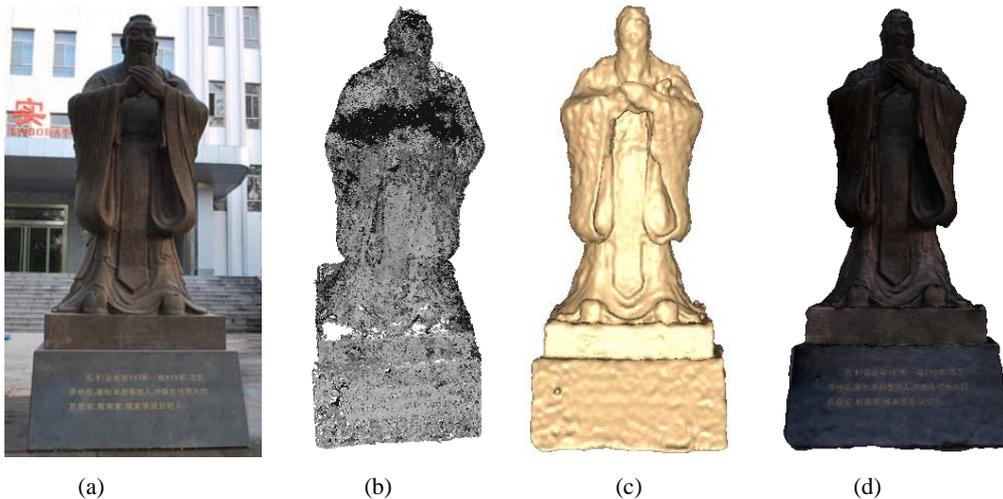

(a)        (b)        (c)        (d)

Figure 6: Reconstructions of the Confucius statue. (a) A photograph of the statue. (b) The oriented points. (c) The reconstructed surface and (d) a textured view.

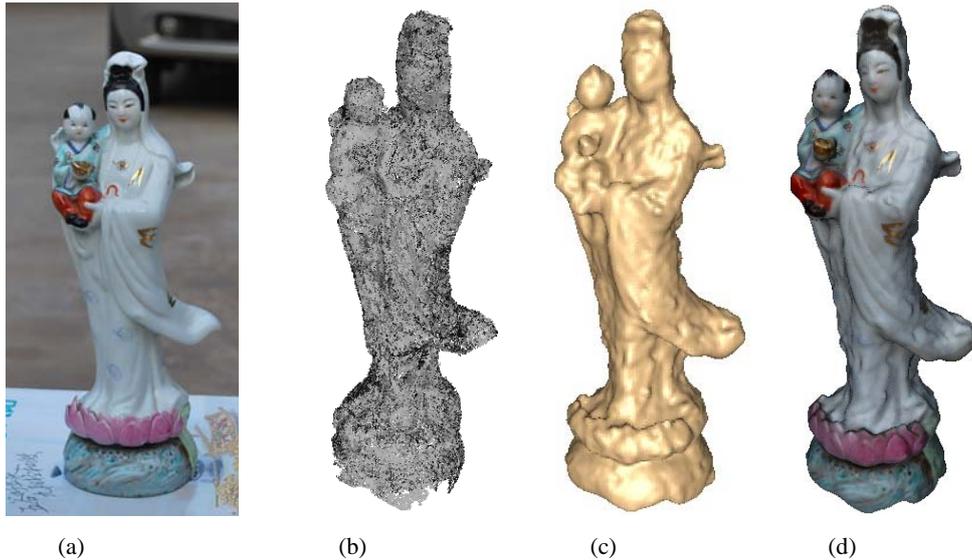

| (a) | (b) | (c) | (d) |

Figure 7: Reconstructions of the Goddess of Mercy statue. (a) A photograph of the statue. (b) The oriented points. (c) The reconstructed surface and (d) a textured view.

# 6 Conclusion

We have presented a method to reconstruct surfaces with higher-order smoothness from 3D measurements. The method is robust to noise, large holes, non-uniform sampling density and very coarse orientation information.

There are several future works to pursue. Because of discretizing on a regular 3D grid, the problem becomes impractical for much fine-detailed reconstruction. We intend to extend this work by an adaptive structure. Future work will also include speed optimization using parallel computing.

## Acknowledgments

The authors would like to express their thanks to Victor Lempitsky and Michael Misha Kazhdan for fruitful discussions. The authors would also like to thank Rasmus R. Paulsen, Noah Snavely and Yasutaka Furukawa for making their algorithms publicly available. This work was supported by the Key Project in the National Science & Technology Pillar Program of China (Grant No.2008BAH29B02), Shandong Natural Science Foundation of China (Grant No. ZR2010FM046) and projects of the Ministry of Education of the Czech Republic (No. 2C06002 and ME10060).